\documentclass[doublecol,linenumbers]{epl2}
\usepackage{graphicx}
\usepackage{amsmath}
\usepackage{subfigure}

\title{Ordered amorphous spin system}

\author{G. Wolff \and D. Levine}
\shortauthor{G. Wolff \etal}

\institute{Department of Physics, Technion, Haifa 32000, Israel}

\pacs{75.10.Nr}{Spin-glass models}
\pacs{61.43.Fs}{Structure of glasses}
\pacs{75.50.Lk}{Glasses magnetic materials}

\abstract{
A solid is typically deemed amorphous when there are no Bragg peaks in its diffraction pattern.  We discuss a two dimensional configuration of Ising spins with an autocorrelation function which vanishes at all nonzero distances, so that its scattering pattern is flat. This configuration is a ground state of a Hamiltonian with deterministic, translationally-invariant and finite range interactions. Despite ostensibly being amorphous, this configuration has perfect underlying  order.  The finite temperature behavior of this model exhibits ordering transitions at successively larger length scales.}

\begin{document}

\maketitle

The question of whether a system is considered ordered is an evolving subject.  Before 1984, order was equated with crystallinity; thereafter, quasicrystals---aperiodic solids with perfect long-range translational order---were shown to be possible\cite{LevineSteinhardt84}.  All other solids were regarded as glassy, or amorphous.  It is natural to ask whether any other types of organization may be perfectly ordered, while being neither crystalline nor quasicrystalline.

Recently, a new criterion for the definition and quantification of long-range spatial order based on ``patch entropy'' was proposed\cite{KurchanLevine11}.  This definition subsumes crystals and quasicrystals, and includes many other systems which would have been classified as disordered due to their scattering spectrum.  One example of this is the Rudin-Shapiro sequence\cite{AlloucheShallit03}, which may be thought of as a 1D configuration of Ising spins, which has an absolutely continuous flat Fourier spectrum. Given that such configurations are mathematically possible, the next important question is whether they are physically relevant.  One way to address this is to ask whether the configuration can be the ground state of some reasonable Hamiltonian.

In this Letter we present a Wang tiling\cite{GrunbaumShephrad86} which is a 2D generalization of the Rudin-Shapiro sequence.  One interpretation of this tiling is as a configuration of Ising spins ($\sigma = \pm 1$) on a square lattice; as such it has been studied in the context of Hadamard matrices\cite{Frank03}.   Although its autocorrelation function is identically zero for all nonzero distances, it is perfectly ordered in the sense of Reference \cite{KurchanLevine11}. Moreover, we shall show that this tiling is the ground state of a short-range (nearest neighbor) Hamiltonian, which we use to study its behavior at finite temperature. Lastly, we note that although there is a degeneracy of ground states, it is subextensive, in the sense of Reference \cite{KurchanLevine11}.

Although this system may be thought of as somewhat artificial, the fact that it minimizes the energy resulting from a local interaction lends credence to the suggestion that physical systems may realize ordered states which are neither crystalline nor quasicrystalline at low enough temperatures.  Such states would not be distinguished by Bragg peaks in their scattering function, and thus might be incorrectly categorized as disordered.

To develop the tiling, we first consider several related polynomials\footnote{These are multivariable generalizations of the Rudin-Shapiro polynomials\cite{Shapiro51}}.  Denoting the spin at the (square) lattice vertex $k,l$ by $\sigma_{k,l}$, we define the polynomial $P(x,y)$ by
\begin{equation*}
P(x,y)= \sum_{k,l}\sigma_{k,l} \; x^k y^l 
\end{equation*}
where $x, y$ are complex variables.  Note that if $x,y$ are defined on the unit circle, we may write $x = e^{iq_{x}}$ and $y = e^{iq_{y}}$, which, when substituted into $P(x,y)$, yields the lattice Fourier transform of the $\{ \sigma_{k,l} \}$:
\begin{equation*}
\mathcal{F}[\{\sigma\}] = \sum_{k,l}\sigma_{k,l} \; e^{ikq_x}e^{ilq_y}
\end{equation*}

Now, if the amplitude of the Fourier transform is a constant, independent of $(q_x,q_y)$, the autocorrelation function must vanish except at the origin:
\begin{equation}
\langle \sigma_{k,l} \sigma_{k+\xi,l+\eta} \rangle = \delta_{\xi,0}\delta_{\eta,0}
\label{eq:autocorr}
\end{equation}
Here the average of a quantity $Q_{k,l}$ is given by $\langle Q_{k,l}  \rangle \equiv \frac{1}{N}\sum_{k,l} Q_{k,l}$ as $N\to\infty$. Thus the problem of finding a ``deterministic amorphous'' configuration of spins reduces to finding a deterministic polynomial in the variables $x = e^{iq_{x}}$ and $y=e^{iq_{y}}$ which has a constant modulus.

Let us now consider  polynomials $P_{n} \equiv P_n (x,y)$, and define them recursively using three auxiliary polynomials, $Q_n,R_n,S_n$ (where the dependence on $x,y$ has been omitted for ease in reading) :
\begin{equation}
\begin{split}
	P_{n+1}&=P_n+x^{2^n}Q_n+y^{2^n}R_n+(xy)^{2^n}S_n\\
	Q_{n+1}&=P_n+x^{2^n}Q_n-y^{2^n}R_n-(xy)^{2^n}S_n\\
	R_{n+1}&=P_n-x^{2^n}Q_n+y^{2^n}R_n-(xy)^{2^n}S_n\\
	S_{n+1}&=P_n-x^{2^n}Q_n-y^{2^n}R_n+(xy)^{2^n}S_n\\
	P_0&=Q_0=R_0=S_0=1
\end{split}\label{eq:polynomial_definition}
\end{equation}
$P_{n},Q_{n},R_{n},S_{n}$ are all polynomials of degree $2^{n+1}-2$ and their coefficients take the values $\pm1$.  In particular, $P_{n}(x,y) = \sum_{k,l=0}^{2^n-1} \sigma_{kl} \; x^{k}y^{l}$, where $\sigma_{kl} = \pm 1$.  Thus we may consider $\sigma_{kl}$ to be a configuration of Ising spins on a square lattice of linear size $2^{n}$, with the total number of spins being $N=4^n$.

If $x$ and $y$ have unit modulus, we have that
\begin{equation*}
\begin{split}
	|P_{n+1}+Q_{n+1}+R_{n+1}+S_{n+1}|^2=16|P_n|^2\\
	|P_{n+1}+Q_{n+1}-R_{n+1}-S_{n+1}|^2=16|Q_n|^2\\
	|P_{n+1}-Q_{n+1}+R_{n+1}-S_{n+1}|^2=16|R_n|^2\\
	|P_{n+1}-Q_{n+1}-R_{n+1}+S_{n+1}|^2=16|S_n|^2
\end{split}\end{equation*}
Summing these, we get $W_{n+1}=4W_n$, where $W_n\equiv|P_n|^2+|Q_n|^2+|R_n|^2+|S_n|^2$. With the initial condition $W_0=4$, we have $W_n=4^{n+1}$, so $|P_n| \le 2^{n+1} \sim \sqrt{N}$. This implies an absolutely continuous Fourier spectrum\cite{square-root-property}.  In fact, as shown by Frank\cite{Frank03}, as $N\to \infty$, the Fourier spectrum is \textit{flat}, i.e. $|P_n| $ is constant, independent of $(q_x,q_y)$, with $|P_n|\to\sqrt{N}$.

The terms in $P_n$, which are all of the form $\pm x^ky^l$, can be organized in the form of a $2^n\times2^n$ matrix $\mathcal{P}_n$ such that if $\sigma_{k,l}x^ky^l\in P_n(x,y)$ then $(\mathcal{P}_n)_{l,k}=\sigma_{k,l}$. In this manner we also define matrices for the three auxilliary polynomials, and denote them $\mathcal{Q}_n,\mathcal{R}_n,\mathcal{S}_n$. This allows us to write the recursive definition~\eqref{eq:polynomial_definition} as
\begin{equation}
\begin{matrix}
\mathcal{P}_{n+1}= \begin{pmatrix} \mathcal{P}_n & \mathcal{Q}_n \\ \mathcal{R}_n & \mathcal{S}_n \end{pmatrix} & \mathcal{Q}_{n+1}= \begin{pmatrix}\mathcal{P}_n & \mathcal{Q}_n \\ -\mathcal{R}_n & -\mathcal{S}_n\end{pmatrix}\\
\mathcal{R}_{n+1}= \begin{pmatrix} \mathcal{P}_n & -\mathcal{Q}_n \\ \mathcal{R}_n & -\mathcal{S}_n \end{pmatrix} & \mathcal{S}_{n+1}= \begin{pmatrix}\mathcal{P}_n & -\mathcal{Q}_n \\ -\mathcal{R}_n & \mathcal{S}_n\end{pmatrix}
\end{matrix}
\label{eq:matrix_recursion} \end{equation}
with $\mathcal{P}_0=\mathcal{Q}_0=\mathcal{R}_0=\mathcal{S}_0=1$.

\begin{figure}
\includegraphics[width=\columnwidth]{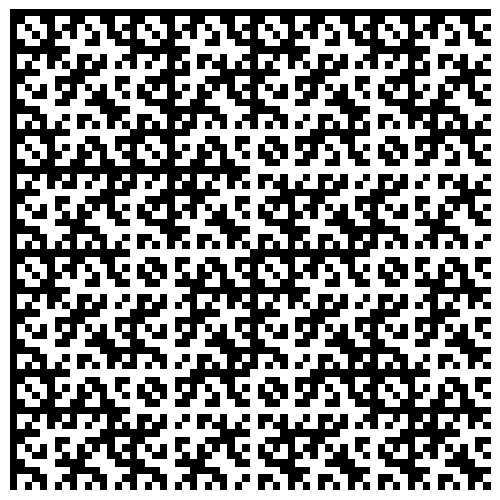}
\caption{Part of the configuration of Ising spins that is a generalization of the Rudin-Shapiro sequence. Each black square corresponds to a spin $\sigma=1$, and each white square corresponds to a spin $\sigma=-1$. Here $0\leq k,l<64$. \label{fig:2D_configuration}}
\end{figure}

In each iteration, the linear size of the matrices $\mathcal{P}_n,\mathcal{Q}_n,\mathcal{R}_n,\mathcal{S}_n$ is doubled. After $n$ iterations each matrix has $N=4^n$ entries, which are $\pm 1$. We regard the matrix $\mathcal{P}_{n}$ as a configuration of Ising spins on a square lattice; Fig.~\ref{fig:2D_configuration} shows a configuration of $64^{2}$ spins ($n=6$).

A glance at Fig.~\ref{fig:2D_configuration} shows that unlike a random configuration, certain patches appear regularly; nonetheless, the two-point correlation function vanishes when all the sites are summed over.  Additionally, as must be the case, there are higher order correlation functions which do not vanish; in our system, for example, certain four-point correlations are non-zero.  We also note that a consequence of~\eqref{eq:autocorr} is that the Parisi overlap function\cite{Mezard87} $P(q)$, measuring the distribution of the overlap, $q$, of patches of the configuration\cite{vanEnter92}, tends to a delta function centered on $q=0$, as it does for a random configuration.

\section{Substitution Rules} Eq.~\eqref{eq:matrix_recursion} implies a recursive construction rule with which we may construct a Wang tiling, which is a tiling of the plane with square tiles.  By looking at the matrix $\mathcal{P}_{n+1}$, we note that a configuration at generation $n$ can be subdivided into four quadrants; each representing a configuration at generation $n-1$. These quadrants may again be subdivided. This iterative procedure ``bottoms out'' at generation, say\footnote{We may choose $n=0$ but then the substitution rules, which would now take a single spin into four, would depend on the spin's location}, $n=1$. The configuration $\mathcal{P}_n$ is now subdivided into $2\times 2$ blocks of the form $\pm \mathcal{P}_1,\pm \mathcal{Q}_1,\pm \mathcal{R}_1,\pm \mathcal{S}_1$. If we simultaneously replace these $2\times 2$ blocks by $4\times 4$ blocks as prescribed by Eq.~\eqref{eq:matrix_recursion} we transform $\mathcal{P}_n$ into $\mathcal{P}_{n+1}$. This procedure allows us to forgo configurations $\pm \mathcal{Q}_n,\pm \mathcal{R}_n,\pm \mathcal{S}_n$ for $n>1$ and inflate the configuration $\mathcal{P}_n\to\mathcal{P}_{n+1}$ directly.

\begin{figure}[h]
\includegraphics[width=\columnwidth]{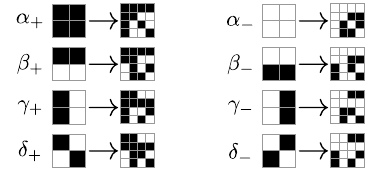}
\caption{The eight basic tiles in the `semaphore' representation, and their substitution rules. A black square represents a spin $\sigma=1$ and a white square represents a spin $\sigma=-1$. The eight tiles are defined in Eq.  \eqref{eq:8_letter_alphabet} and their substitution rules in \eqref{eq:substitution_rules}. The substitutions generate the configuration $\mathcal{P}$ when iterated on a $2\times 2$ block of all black squares. 
\label{fig:2D_substitutions}}
\end{figure}

Let us now denote each of the eight $2\times 2$ blocks by a letter from the alphabet $\{\alpha_{+},\alpha_{-},\beta_{+},\beta_{-},\gamma_{+},\gamma_{-},\delta_{+},\delta_{-}\}$ as follows
\begin{equation}\label{eq:8_letter_alphabet}
\begin{split}
	\alpha_{+}=-\alpha_{-}=\mathcal{P}_1=\begin{array}{cc}+1&+1\\+1&+1\end{array},\\
	\beta_{+}=-\beta_{-}=\mathcal{Q}_1=\begin{array}{cc}+1&+1\\-1&-1\end{array},\\
	\gamma_{+}=-\gamma_{-}=\mathcal{R}_1=\begin{array}{cc}+1&-1\\+1&-1\end{array},\\
	\delta_{+}=-\delta_{-}=\mathcal{S}_1=\begin{array}{cc}+1&-1\\-1&+1\end{array},
\end{split}
\end{equation}
Each of these letters will represent a set of tiles in the Wang tiling. For ease in visualization, we will represent each of the tiles as a 
$2\times2$ `semaphore' block, with 1 represented by a black square and $-1$ by a white square, as shown in Figure \ref{fig:2D_substitutions}.

By Eq.~\eqref{eq:matrix_recursion}, each of the above $2\times 2$ blocks is to be substituted by a $4\times 4$ block in the following manner
\begin{equation}\label{eq:substitution_rules}\begin{split}
	\alpha_{+} &\to \begin{array}{cc}\alpha_{+} &\beta_{+}\\ \gamma_{+}&\delta_{+}\end{array}			\qquad
	\alpha_{-} \to \begin{array}{cc}\alpha_{-} &\beta_{-}\\ \gamma_{-} &\delta_{-}\end{array}	\\
	\beta_{+}&\to \begin{array}{cc}\alpha_{+} &\beta_{+}\\ \gamma_{-} &\delta_{-}\end{array}		\qquad
	\beta_{-}\to \begin{array}{cc}\alpha_{-} &\beta_{-}\\\gamma_{+}&\delta_{+}\end{array}			\\
	\gamma_{+}&\to \begin{array}{cc}\alpha_{+} &\beta_{-}\\\gamma_{+}&\delta_{-}\end{array}			\qquad
	\gamma_{-} \to \begin{array}{cc}\alpha_{-} &\beta_{+}\\ \gamma_{-} &\delta_{+}\end{array}\\
	\delta_{+}&\to \begin{array}{cc}\alpha_{+} &\beta_{-}\\ \gamma_{-} &\delta_{+}\end{array}		\qquad
	\delta_{-}\to \begin{array}{cc}\alpha_{-} &\beta_{+}\\\gamma_{+}&\delta_{-}\end{array}			.
\end{split}\end{equation}
These substitution rules are shown in Figure~\ref{fig:2D_substitutions}. 

The substitution rules obey two important properties: (i) each of the pairs $(\alpha_{+},\alpha_{-})$, $(\beta_{+},\beta_{-})$, $(\gamma_{+},\gamma_{-})$, and $(\delta_{+},\delta_{-})$ occupy a specific quadrant on the rhs of each substitution rule, (ii) in each pair, one partner is the negative of the other, and its substitution rule is the negative of its partner's substitution rule.  By repeatedly applying these rules, we can generate a tiling of arbitrary size.

\section{Matching Rules}By a theorem of Mozes\cite{Mozes89}, configurations emerging from substitution rules such as those in Eq.~\eqref{eq:substitution_rules} may be enforced by a finite set of Wang tiles---each one with different markings on the edges. The plane is to be tiled by copies of these tiles\footnote{Only translations of the Wang tiles are allowed; rotations and reflections are forbidden.} such that two tiles may abut if and only if the markings on their common edge match---these are the so-called {\it matching rules}.  We note that although the configuration above is generated substitutionally from eight basic tiles, there will be more than eight Wang tiles in terms of their markings.

The matching rules may be interpreted as interactions for purposes of physics:  By associating an energy penalty $+J$ for each edge with mismatched markings, we define a Hamiltonian
\begin{equation}\label{eq:tiling_Hamiltonian}
\mathcal{H}=J\sum_{x,y}2-\delta\left(T_{x,y}^N-T_{x,y+1}^S\right)-\delta\left(T_{x,y}^E-T_{x+1,y}^W\right)
\end{equation}
where $\delta(0) = 1$ and $\delta(x\ne0) = 0$ and where $T$ is a tile with edge markings $T^N,T^S,T^E,T^W$.
The ground state of this Hamiltonian has zero energy and corresponds to a tiling in which there are no mismatches. 
 
We note that this system may be regarded as a generalized Potts model.  Let $t$ be a Potts spin, taking $M$ possible values, where $M$ is the number of distinct Wang tiles. We denote these values by $t=1\ldots M$. Since two nearest-neighbor tiles have an interaction energy ($J$ or $0$), the Hamiltonian~\eqref{eq:tiling_Hamiltonian} can be written as
\begin{equation}\label{eq:Potts}
\mathcal{H}=J\sum_{x,y}Q^{NS}(t_{x,y},t_{x,y+1})+Q^{EW}(t_{x,y},t_{x+1,y})
\end{equation}
where $Q^{NS}$ and $Q^{EW}$ are two $M\times M$ matrices whose entries are $0$ and $1$.  In a regular Potts model each $Q$ matrix would have zeros on the diagonal and ones off the diagonal; here the matrices reflect the matching rules between pairs of tiles in their various juxtapositions.

In order to fully define the system, we must construct the matrices $Q$ used in Eq.~\eqref{eq:Potts}, or equivalently, construct a set of Wang tiles (including matching rules) which allow our configurations as the only legal tilings of the plane. We have constructed such a set by the two-step procedure described below.

\textbf{Step 1}. Apply Mozes' construction~\cite{Mozes89} to our tiling.  In this construction there are two fundamental types of tiles. The first type corresponds to the substitutional (``letter'')  tiles (in our case $\alpha_{+},\alpha_{-},\beta_{+},\beta_{-},\gamma_{+},\gamma_{-},\delta_{+},\delta_{-}$). These tiles are constrained to occupy the sites of a diluted lattice of double the spacing of the physical lattice (\textit{e.g.,} sites where $x$ and $y$ are both even). The second type corresponds to tiles whose function is to synchronize~\cite{Mozes89} the letter tiles so that those will order into the tiling shown in Figure~\ref{fig:2D_configuration}. The ``synchronizer'' tiles are not allowed to occupy the sites of the diluted lattice. This construction leads to a tile set consisting of 116 tiles\footnote{See \texttt{tx.technion.ac.il/\~{}gilwolff/116tiles.pdf} for the set of tiles}, of which 8 are letter tiles and 108 are synchronizer tiles. The Hamiltonian is not yet that of Eq.~\eqref{eq:tiling_Hamiltonian} because different tiles are restricted to different sites of the lattice.

\textbf{Step 2}. Define ``supertiles'', each a plaquette consisting of one letter tile and the three synchronizer tiles lying to its right, bottom, and bottom-right. The edge-markings of the supertiles are the superposition of the edge-markings of its constituent tiles. A careful combinatorical counting shows that this step generates 400 supertiles, for which the Hamiltonian is that of Eq.~\eqref{eq:tiling_Hamiltonian}.

The 400 tiles differ by their edge markings but represent variations of the letter tiles $\{\alpha_{+},\alpha_{-},\beta_{+},\beta_{-},\gamma_{+},\gamma_{-},\delta_{+},\delta_{-}\}$. Returning to \eqref{eq:Potts}, we note that half of the tiles are variations of $\{\alpha_{+},\beta_{+},\gamma_{+},\delta_{+}\}$ while the other half are variations of $\{\alpha_{-},\beta_{-},\gamma_{-},\delta_{-}\}$. Therefore eq.~\eqref{eq:Potts} can be rewritten as 
\begin{equation*}
\mathcal{H}=J\sum_{x,y}Q^{NS}(\sigma^\mu_{x,y} \, ,\sigma^\nu_{x,y+1})+Q^{EW}(\sigma^\mu_{x,y} \, ,\sigma^\nu_{x+1,y})
\end{equation*}
where the indices $\mu,\nu$ run over the 200 variations of $\sigma=\pm1$. 

\begin{figure*}
\subfigure{\includegraphics[height=7.27cm]{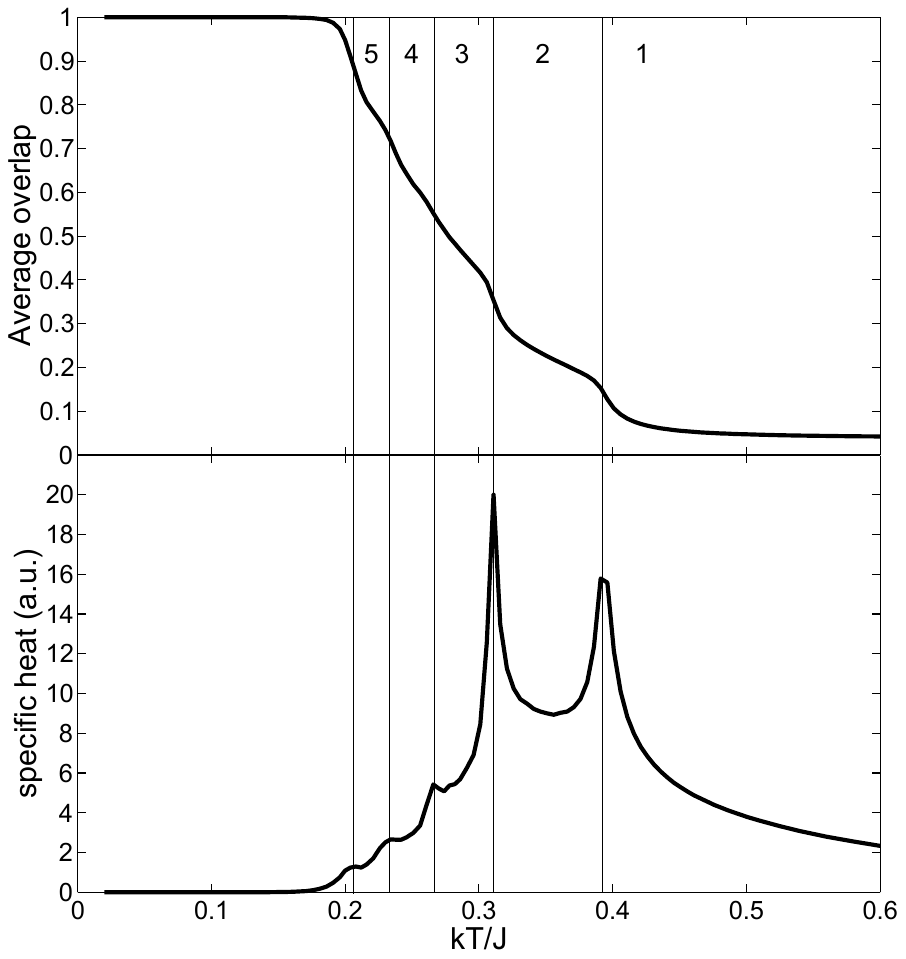}}
\subfigure{\includegraphics[height=7.2cm]{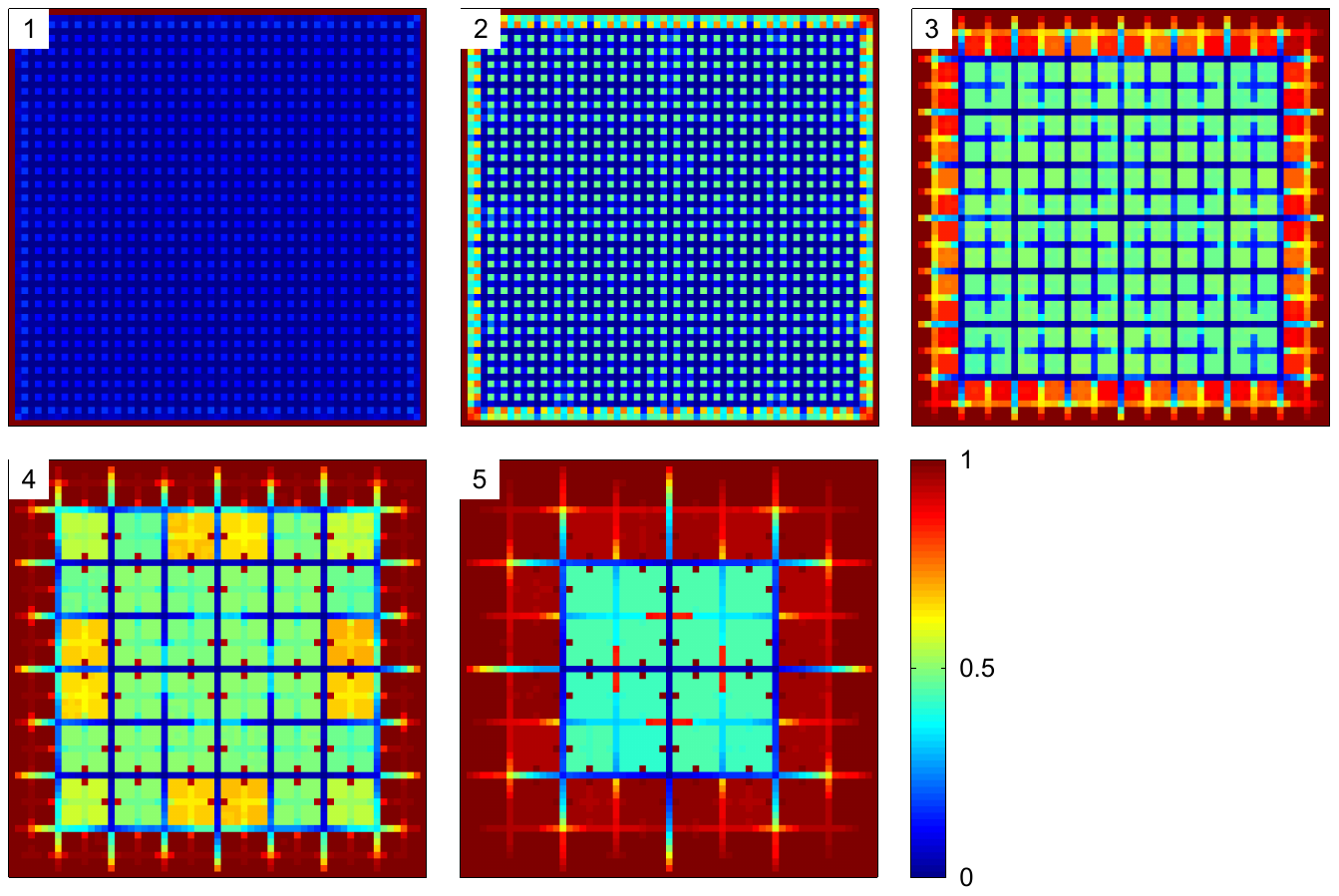}}
\caption{(color online) Finite $T$ behavior of the restricted Wang tiling. Left: Specific heat and average overlap as a function of temperature ($J$ is the unit of energy associated with a pair of mismatched edges in the Wang tiling). Right: visualization of the phases via the on-site overlap between them and the ground state. \label{fig:finite_T}}
\end{figure*}

\section{Finite $T$ Behavior}
To investigate behavior at finite temperature, we performed Monte-Carlo (MC) simulations on the 116 tile model.  The price to pay for using this (comparatively) small tile set is that, as stated above, in this model, different tiles are restricted to different sites of the lattice.  We note that a preliminary investigation of the (unrestricted) 400 tile model shows the qualitative behavior of the two models to be the same, so we have concentrated our efforts in the 116 tile model to reduce computational complexity.

We performed simulations on systems of linear length $L=15,31,63,127$ with closed boundary conditions, and measured the specific heat and the average overlap\cite{LeuzziParisi00} between a configuration at temperature $T$ and the ground state. Results for the system of length $L=63$ are presented in the left of Fig.~\ref{fig:finite_T}, and indicate that the system undergoes a sequence of phase transitions, as the system orders on successively larger length scales. The average overlaps of five of these phases with the ground state are shown on the right of Fig.~\ref{fig:finite_T}.  In all cases, tiles on the border of the system are fixed, so that their overlap is always unity.

Phase 1 is the high-temperature disordered phase. Sites with both coordinates even---where the eight letter tiles are restricted to lie---have an average overlap $1/8$; the remaining sites have average overlap $1/108$ \footnote{This is the reason for the two shades of blue.}.  In phase 2, a periodic array of lattice constant 2, corresponding to $x$ and $y$ both even, has set into place---the identity ($\alpha,\beta,\gamma,\delta$) of each letter tile has frozen into the arrangement
\begin{equation*}
\begin{array}{ccccccc}
& | &  & | &  & |\\
 - & \alpha & - & \beta & - & \alpha & -\\
 & | &  & | &  & |\\
 - & \gamma & - & \delta & - & \gamma & -\\
 & | &  & | &  & |\\
 - & \alpha & - & \beta & - & \alpha & -\\
  & | &  & | &  & |
\end{array}
\end{equation*}
but the sign of each letter tile (the subscript, plus or minus) is still random. This gives an average overlap 1/2  for these sites, indicated by green in Fig.~\ref{fig:finite_T}.

In Phase 3 each $3\times 3$ block (which contains 4 letter tiles and 5 synchronizer tiles) is internally arranged, meaning that the signs of the letter tiles in that block conform to one of the substitution rules in \eqref{eq:substitution_rules}. They are also externally arranged in the sense that if one block's letter tiles are are taken from the substitution rule of $\alpha_+$ or $\alpha_-$, then the horizontally adjacent block's letter tiles are taken from the substitution rule of $\beta_+$ or $\beta_-$, and the vertically adjacent block's letter tiles are taken from the substitution rule of $\gamma_+$ or $\gamma_-$. Each $3 \times 3$ block therefore has average overlap 1/2.  In phase 4 this again happens for blocks of size $7\times 7$ (which contain 16 letter tiles and 33 synchronizer tiles). They are internally arranged, but externally arranged only up to their sign. In phase 5 the size of the blocks increases to $15\times 15$.  This behavior is similar to results on hierarchical tilings~\cite{ByingtonSocolar12,Miekisz90}, and we expect there to be ordering transitions on successively larger scales in the thermodynamic limit.

The model we have studied raises the question of whether there exist materials which have been classified as amorphous which are nevertheless actually ordered.  This type of order would not reveal itself in conventional scattering experiments, and it might not be obvious which correlations are present, but it would, however, be manifested in the behavior of the patch entropy, which does not require {\it a-priori} knowledge of the nature of the ordering.

\acknowledgments
We would like to thank J-P. Allouche, M. Baake, A. van Enter, U. Grimm, D. Hexner, J. Kurchan, N. Nikola, and J. Socolar for interesting discussions.  DL thanks the US-Israel Binational Science Foundation (grant 2008483) and the Israel Science Foundation (grant 1254/12) for support.

\bibliographystyle{eplbib}
\bibliography{ref}

\end{document}


\title{Supplemental material for paper ``ordered amorphous spin system"}
\maketitle

Here we give the set of 116 Wang tiles which enforce an ``amorphous'' tiling of the plane. The markings $\oplus,\ominus,b,t,r,\ell,s,m,\boxplus,\boxminus,B,T,R,L,S,M,h,H,v,V$ must match on the shared edge between abutting tiles.

\textbf{The 8 Letter Tiles}. These are tiles that correspond to the substitutional letters $\alpha_{+},\alpha_{-},\beta_{+},\beta_{-},\gamma_{+},\gamma_{-},\delta_{+},\delta_{-}$. They are restricted to occupy the sites of a diluted lattice of double the spacing of the physical lattice, \emph{e.g.} sites where $x,y$ are both even.
\begin{figure}[!h]
\includegraphics[width=0.6\columnwidth]{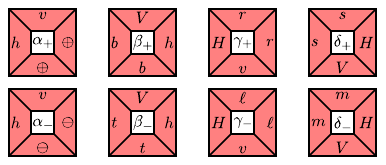}
\end{figure}

\textbf{The 108 Synchronizer tiles}. These tiles are restricted to occupy the remaining $\sfrac{3}{4}$ sites of the lattice. The 108 tiles are arranged in groups; the tiles in each group serve a common function in synchronizing the 8 letter tiles. The coloring of the groups merely serves to clarify the example on the following page

\begin{figure}[!h]
\includegraphics[width=0.6\columnwidth]{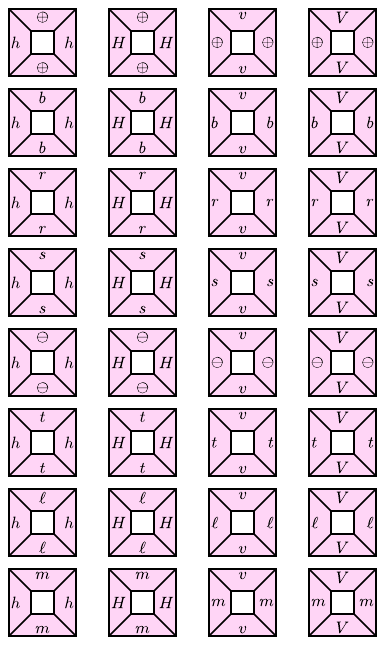}
\end{figure}

\begin{figure}[!h]
\includegraphics[width=0.6\columnwidth]{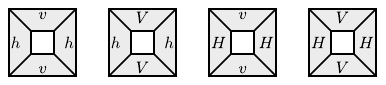}
\end{figure}

\pagebreak

\begin{figure}[!h]
\includegraphics[width=0.6\columnwidth]{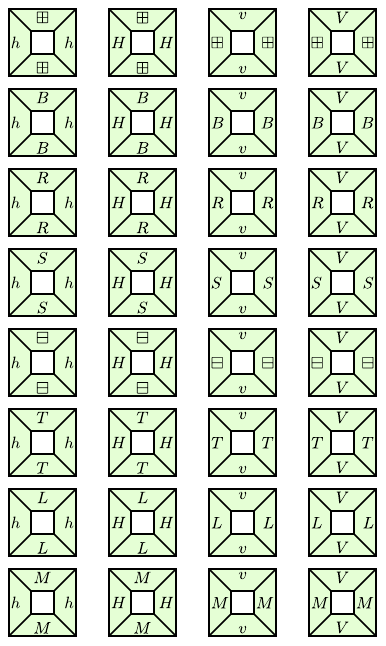}
\end{figure}
\nopagebreak
\begin{figure}[!h]
\includegraphics[width=0.6\columnwidth]{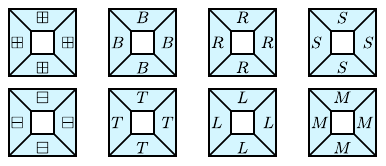}
\end{figure}
\nopagebreak
\begin{figure}[!h]
\includegraphics[width=0.6\columnwidth]{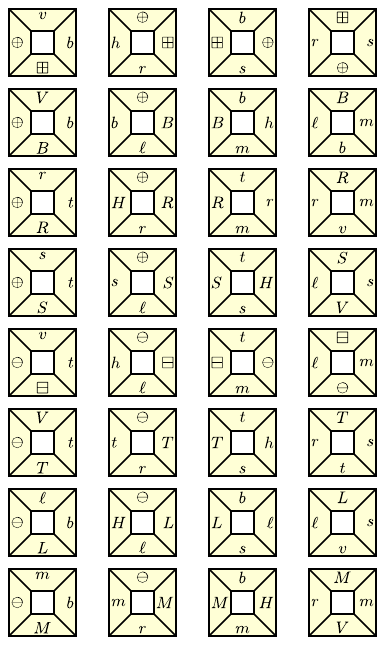}
\end{figure}

\pagebreak

\textbf{An example}. The following is an example of a legal tiling. It is a portion of the infinite tiling which obeys the matching rules
\nopagebreak
\begin{figure}[!h]
\includegraphics[width=\columnwidth]{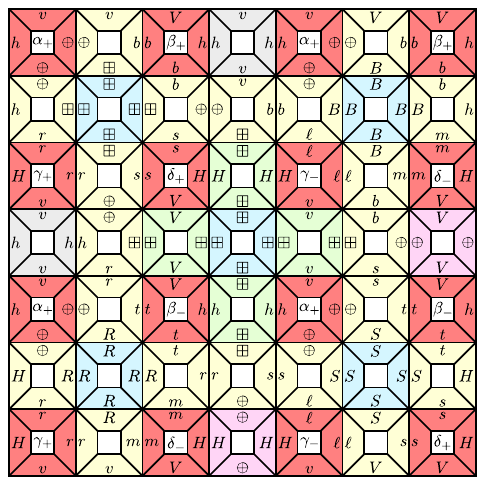}
\end{figure}